\documentstyle[12pt]{article}

\catcode`@=11 \@addtoreset{equation}{section} \catcode`@=12

\newcommand{\alphas}{\alpha_{\scriptscriptstyle S}}
\newcommand{\jpsi}{J/\psi}
\newcommand{\ptr}{p_{\scriptscriptstyle T}}

\newcommand{\mochij}{\left\langle{\cal O}^{\chi_J}_8({}^3S_1)\right\rangle}
\newcommand{\mochiz}{\left\langle{\cal O}^{\chi_0}_8({}^3S_1)\right\rangle}
\newcommand{\mochio}{\left\langle{\cal O}^{\chi_1}_8({}^3S_1)\right\rangle}

\newcommand{\mschij}{\left\langle{\cal O}^{\chi_J}_1({}^3P_J)\right\rangle}
\newcommand{\mschiz}{\left\langle{\cal O}^{\chi_0}_1({}^3P_0)\right\rangle}

\newcommand{\mschit}{\left\langle{\cal O}^{\chi_2}_1({}^3P_2)\right\rangle}
\newcommand{\mojpsa}{\left\langle{\cal O}^{J/\psi}_8({}^3S_1)\right\rangle}
\newcommand{\mojpsb}{\left[
          \left\langle{\cal O}^{J/\psi}_8({}^1S_0)\right\rangle
    +{\displaystyle3\over\displaystyle m^2}
          \left\langle{\cal O}^{J/\psi}_8({}^3P_0)\right\rangle
    +{\displaystyle4\over\displaystyle5m^2}
          \left\langle{\cal O}^{J/\psi}_8({}^3P_2)\right\rangle\right]}

\begin{document}
\begin{titlepage}
\begin{flushright}\vbox{\begin{tabular}{c}
           TIFR/TH/96-04\\
           January 25, 1996\\
           hep-ph/9601349
\end{tabular}}\end{flushright}
\begin{center}
   {\large \bf  Colour-octet Contributions to $J/\psi$\\
                Hadroproduction at Fixed Target Energies.}
\end{center}
\bigskip
\begin{center}
   {Sourendu Gupta and K.\ Sridhar.\\
    Theory Group, Tata Institute of Fundamental Research,\\
    Homi Bhabha Road, Bombay 400005, India.}
\end{center}
\bigskip
\begin{abstract}
We investigate the integrated cross section for forward $\jpsi$ production
in proton-proton and pion-proton collisions at centre of mass energies
upto 60 GeV. We find that colour-octet contributions to the cross section
are crucially important, and their inclusion produces a very good description
of the data. Since the values of non-perturbative matrix elements are fixed
by other experiments, these predictions are parameter-free.
\end{abstract}
\end{titlepage}

The production of bound states of heavy quarks has been
theoretically studied in terms of the colour-singlet model
\cite{sing}. In this approach, the dominant
contribution to quarkonium production is expected to come
from quark-antiquark or gluon-gluon fusion, leading to the
formation of a heavy-quark pair in a colour-singlet state
with the correct spin, parity and charge-conjugation assignments
projected out. By using radial wave-functions or their derivatives
derived from other measurements as inputs, the colour-singlet model
makes definite predictions for the production cross sections of
specific quarkonium resonances.

While this model gives a reasonable
description of large-$\ptr$ $\jpsi$ production at ISR energies,
it completely fails to explain the $\ptr$ integrated cross sections
at fixed-target and ISR energies. This failure is easily traced
back to the model assumptions. The requirement that a colour-singlet
state with the correct quantum numbers be produced at the hard
vertex imposes rather stringent constraints (coming from $C$-invariance)
and rules out the zero-$p_T$ production of the $S$-state quarkonia
at lowest order. Thus, to leading order, the inclusive cross section
for $\jpsi$ production is obtained only from direct production of
$\chi_c$ followed by their decay to $\jpsi$.

Recent improvements in the theory of quarkonium production have
been provoked by the anomalously large cross section for $\jpsi$
production at large $\ptr$ measured recently in the CDF experiment
at the Tevatron \cite{cdf}. For values of $\ptr$ much larger than the
charm quark mass, fragmentation of gluons and charm quarks become
dominant \cite{bryu}. The CDF data has been successfully explained
by taking into account both the fusion and fragmentation contributions
\cite{jpsi}.

The analyses of the CDF data have brought out another limitation
of the colour-singlet approach, {\sl i.e.\/}, the relative velocity,
$v$, between the heavy quarks in the bound state is ignored. However,
$v$ is often not negligible in quarkonium states and corrections
of order $v$ need to be taken into account. Starting from a 
non-relativistic QCD Lagrangian, a systematic analysis using the 
factorisation method has been recently carried out \cite{nrqcd}. In
this formulation, the quarkonium wave-function admits of a systematic
expansion in powers of $v$ in terms of Fock-space components.
For example, the wave-functions for the $P$-state charmonia have the
conventional colour-singlet $P$-state component at leading order, but
there exist additional contributions at non-leading order in $v$,
involving octet $S$-state components; {\sl i.e.\/}, 
\begin{equation}
  \vert \chi_J \rangle = O(1) \vert Q\bar Q \lbrack {}^3P_J^{(1)} \rbrack
      \rangle + O(v) \vert Q\bar Q \lbrack {}^3S_1^{(8)} \rbrack g
      \rangle + \ldots
\label{e1}\end{equation}
While the octet components are necessary for a consistent perturbative
description of the $P$-states \cite{bbl2}, for $S$-states it is 
required in order to explain the anomalously large $\psi'$ \cite{brfl}
and direct $\jpsi$ production at the Tevatron \cite{cgmp}.

As a consequence of factorisation, each matrix element is split into
two parts. The short-distance parts for the colour-octet process are
calculable, but the long-distance non-perturbative octet matrix elements
(the analogues of the wavefunctions in the colour-singlet model)
are not. They have been obtained by a fit to the Tevatron
data \cite{cgmp}. More recently, the colour-octet processes which 
contribute to photoproduction of $\jpsi$ have been studied
\cite{flem2}. Using both the fixed-target and HERA data on
photoproduction, a different linear combination of the same octet
matrix-elements has been fitted.

The time seems ripe for a reappraisal of the mass of data on hadroproduction
of $\jpsi$ in proton-nucleon and pion-nucleon collisions at centre of mass
energies, $\sqrt S$, upto about 60 GeV (see \cite{schuler} for a compilation
of data). In this paper we carry out this investigation. We find that
the low-energy data on forward ($x_{\scriptscriptstyle F}>0$) integrated
hadroproduction cross sections are well described by the inclusion of octet
components. These are parameter-free predictions, because the values of the
non-perturbative matrix elements are derived from other experiments.

In this model,
a $\jpsi$ can be obtained either directly, or through radiative decays
of $\chi_c$ states. Any of these can be produced through a
colour-singlet channel or through the octet channel. Quantum number
arguments show that the cross sections for direct $\jpsi$ or $\chi_1$
production in the singlet channel are zero (to lowest order in $\alphas$).
As a result, we can write the cross section
\begin{equation}\begin{array}{rl}
   \sigma_{\jpsi}(s)\;=\;\sigma^8_{\jpsi}(s)
     +\sigma^8_{\chi_1}(s) & BR_{\chi_1}
     +\left(\sigma^1_{\chi_0}(s)+\sigma^8_{\chi_0}(s)\right) BR_{\chi_0}\\
     &+\left(\sigma^1_{\chi_2}(s)+\sigma^8_{\chi_2}(s)\right) BR_{\chi_2},
\end{array}\label{sigma}\end{equation}
where $BR_{\chi_J}$ denotes the branching ratio for the decay of
$\chi_{\scriptscriptstyle J}$ into $\jpsi$.
The individual terms are all known. We write down the
forward integrated cross sections below.

The singlet channel cross sections \cite{sing} are given in terms of the
strong coupling $\alphas$ and the charm quark mass $m$ by
\begin{equation}\begin{array}{rl}
     \sigma^1_{\chi_0}\;&=\;G
       {\displaystyle \alphas^2\pi^3\over\displaystyle 48m^7}\mschiz,\\
     \sigma^1_{\chi_2}\;&=\;G
       {\displaystyle \alphas^2\pi^3\over\displaystyle180m^7}\mschit.
\end{array}\label{sing}\end{equation}

The quantity $G$ is defined in terms of the gluon momentum densities in
the projectile (P) and target (T) by
\begin{equation}
   G\;=\;\int_{\sqrt\tau}^1 {dx\over x} g_{\scriptscriptstyle P}(x)
                     g_{\scriptscriptstyle T}(\tau/x).
\label{gluon}\end{equation}
The lower limit of the integral, $\sqrt\tau=2m/\sqrt S$, is written for
the forward cross section. If all values of $x_{\scriptscriptstyle F}$
were to be considered, the limit would have changed to $\tau$.

The two matrix elements are related to the derivative of the wavefunction
at the origin by
\begin{equation}
   \mschij\;=\;{9(2J+1)\over2\pi}|R'(0)|^2.
\label{deriv}\end{equation}

The octet cross section for $\jpsi$ has been calculated recently
\cite{flem1}. It is given by
\begin{equation}\begin{array}{rl}
     \sigma^8_{\jpsi}\;=\;
       G{\displaystyle5\alphas^2\pi^3\over\displaystyle48m^5}&\mojpsb\\
       &+Q{\displaystyle\alphas^2\pi^3\over\displaystyle54m^5}\mojpsa.
\end{array}\label{psi}\end{equation}
The quantity $Q$ is defined in terms of the quark momentum densities in
the projectile and target by
\begin{equation}
   Q\;=\;\sum_f\int_{\sqrt\tau}^1 {dx\over x} q^f_{\scriptscriptstyle P}(x)
                     \bar q^f_{\scriptscriptstyle T}(\tau/x)
        +(P\leftrightarrow T),
\label{quark}\end{equation}
where the sum is over flavours.
The combination of non-perturbative matrix elements in the coefficient of
$G$ is precisely that needed for photoproduction \cite{flem2}.

There are also the octet contributions for the $\chi_c$ states. These can
be easily obtained by noting the following facts. The leading contribution
comes from the ${}^3S_1$ colour-octet states. The gluon fusion contribution
to this channel vanishes \cite{flem1}. Hence, the only cross section required
is
\begin{equation}
   \sigma^8_{\chi_J}\;=\;Q
       {\displaystyle\alphas^2\pi^3\over\displaystyle54m^5}\mochij.
\label{chi}\end{equation}

The singlet matrix elements of eq.\ (\ref{deriv}) are extracted from data
on hadronic decays of the $\chi_c$ states. For a recent compilation see
\cite{mangano}. The colour-octet matrix elements $\mochio$ and $\mojpsa$
have been extracted from the hadroproduction rates at the Tevatron. The
different resonances are identified in these experiments, allowing separate
fits to these matrix elements. We use the tabulation of \cite{cho2}. We
require all the matrix elements $\mochij$. These are related by \cite{nrqcd}
\begin{equation}
   \mochij\;=\;(2J+1)\mochiz.
\label{jscale}\end{equation}
The remaining combination of matrix elements has been extracted from
photoproduction data \cite{flem2}. We use this value
\begin{equation}
   \mojpsb\;=\;0.020\pm0.001\ \ {\rm GeV}^3
\label{value}\end{equation}
in the calculations reported here.

A different linear combination of the same matrix elements enters the
large-$\ptr$ hadroproduction cross sections relevant to the Tevatron
experiments. A consistency check between these two linear combinations
can be made if one assumes a relation between
$\left\langle{\cal O}^{\jpsi}_8({}^3P_2)\right\rangle$ and
$\left\langle{\cal O}^{\jpsi}_8({}^3P_0)\right\rangle$.
Such a relation reduces the number of unknown matrix elements by one
and permits evaluation of the individual matrix elements. With the
assumption that the first matrix element is 5 times the second, the
Tevatron and photoproduction fits together yield a negative value of
$\left\langle{\cal O}^{\jpsi}_8({}^3P_0)\right\rangle$ \cite{flem2}.
This may just indicate that the assumption has to be modified. The
issue is, however, irrelevant to our computation, since we require
precisely the linear combination in eq.\ (\ref{value}).

We use the MRS ${\rm D}-'$ and the GRV LO sets of parton densities for
the proton. Since we are interested in values of $\sqrt S\le60$ GeV,
the lowest values of $x$ entering the parton distributions are
$\sqrt\tau\approx0.05$. For such large values of $x$ the differences
between structure function parametrisations for the proton are rather small.
The pion structure functions are not known as accurately. We have used the
parametrisations SMRS 1, SMRS 3 and GRV. All the
structure functions are taken from the PDFLIB package \cite{pdflib}.

As always in lowest order QCD cross sections, there is a choice of scale
to be considered. In addition, the charm quark mass, $m$ is not precisely
known. From open charm production, the limits are $1.2 {\rm\ GeV}\le m
\le 1.8$ GeV \cite{nason}. Our results are shown for the two choices
$m=1.7$ GeV and a scale of $2m$, as well as $m=1.6$ GeV and a scale of $4m$.

\begin{figure}
\vskip14truecm
\includegraphics{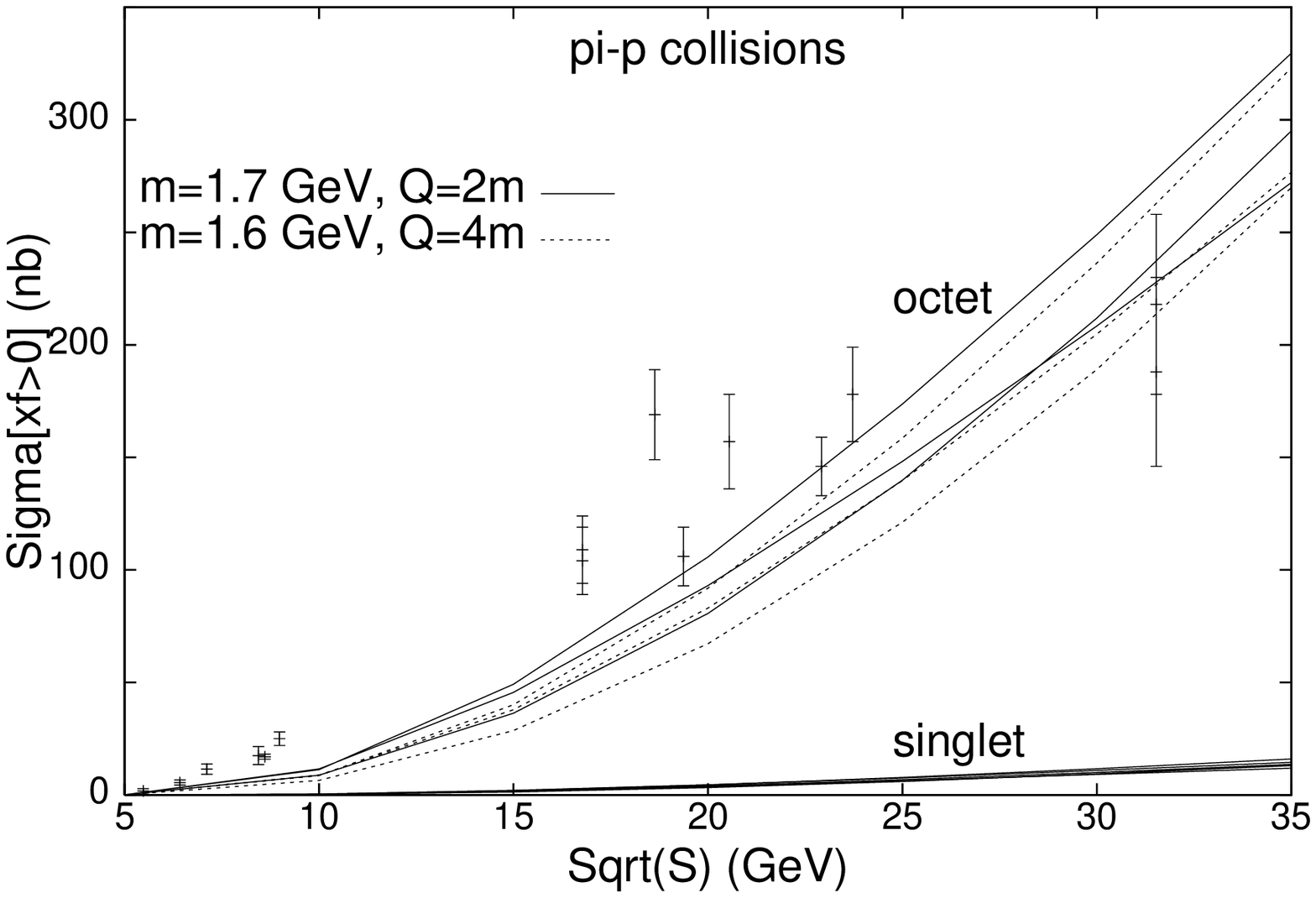}
\includegraphics{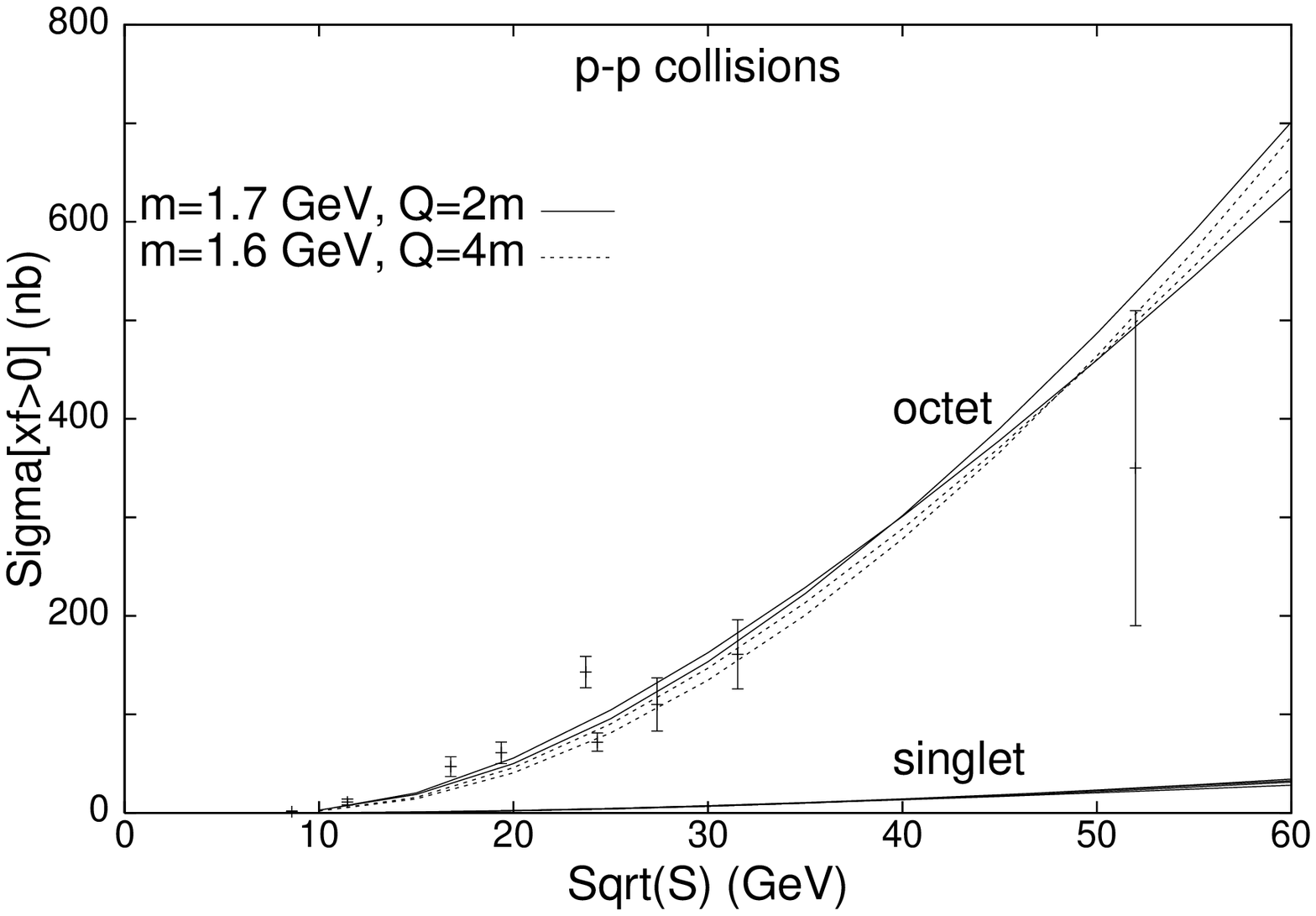}
\caption[dummy]{The colour-octet model predictions for integrated forward
  $\jpsi$ hadroproduction cross sections as a function of the CM energy.
  For $pp$ collisions, the two curves for each choice of $m$ and scale $Q$
  are for the structure functions MRS ${\rm D}-'$ and GRV LO. For $\pi p$
  collisions the three sets of structure functions are
  MRS ${\rm D}-'$ for proton and SMRS 1 for $\pi$, MRS ${\rm D}-'$
  for proton and SMRS 3 for $\pi$, GRV LO for both. Note that the
  colour-singlet model predictions lie far below the data.}
\label{fig}\end{figure}

With the inputs discussed here, we find that the $\sqrt S$ dependence of
the integrated forward $\jpsi$ production rates, for both $pp$ and $\pi p$
collisions, are described rather well by the model (see Figure \ref{fig}).
We would like to emphasise that there are no free parameters in this
calculation.

We find that the total $\jpsi$ hadroproduction rate in $pp$ collision
is dominated by the gluon-gluon fusion part of the colour-octet direct
$\jpsi$ cross section. Although $\chi_0$ is produced copiously, its
contribution to the $\jpsi$ cross section is suppressed by a small
branching ratio. The contribution of the $\chi_1$ is small because the
singlet contribution is zero and the octet is small. The $\chi_2$ state
gives the major contribution among the $P$-channel resonances, but 
nevertheless gives less than 5\% of the inclusive forward cross section.
In $\pi p$ collisions the $P$-channel contributes slightly more to the
total cross section. This is due to the enhancement of the $q\bar q$
initiated octet $\chi$ contributions.

Since the cross section is dominated by the gluon fusion process giving
a colour-octet contribution to direct $\jpsi$, the dominant uncertainty
in the non-perturbative matrix elements comes from the linear combination
in eq.\ (\ref{value}). This gives a 5\% uncertainty in the predictions.
Although the uncertainty in the non-perturbative matrix elements for
colour-octet contributions to the $\chi$ states are larger, they have
a much smaller effect on the result. Similiarly, the assumption stated
in eq.\ (\ref{jscale}) has very little impact on the result.

Larger uncertainties come from the choice of scale in QCD and the charm
quark mass. The scale uncertainty can only be controlled by taking into
account the order $\alphas^3$ contributions to the cross section. When
this is done, it might be possible to control the uncertainty in $m$.

In conclusion, the world data on forward inclusive $\jpsi$ hadroproduction
cross sections are well described by the inclusion of octet contributions.
The calculation is completely parameter-free since
the non-perturbative inputs are fixed from other experiments.
Our results on other aspects
of the phenomenology, such as rapidity and $x_{\scriptscriptstyle F}$
distributions, will be reported elsewhere.

\end{document}